# Real-time observation of electronic, vibrational, and rotational dynamics in nitric oxide with attosecond soft X-ray pulses at 400 eV

NARIYUKI SAITO,[1,*] HIROKI SANNOHE,[1] NOBUHISA ISHII,[2] TERUTO KANAI,[1] NOBUHIRO KOSUGI,[3] YI WU,[4] ANDREW CHEW,[4] SEUNGHWOI HAN,[4] ZENGHU CHANG,[4] AND JIRO ITATANI[1]

[1]*The Institute for Solid State Physics, the University of Tokyo, Kashiwanoha 5-1-5, Kashiwa, Chiba 277-8581, Japan.*
[2]*Kansai Photon Science Institute, National Institutes for Quantum and Radiological Science and Technology, 8-1-7 Umemidai, Kizugawa, Kyoto 619-0215, Japan*
[3]*Institute of Materials Structure Science, High Energy Accelerator Research Organization, 1-1 Oho, Tsukuba, Ibaraki 305-0801, Japan.*
[4]*Institute for the Frontier of Attosecond Science and Technology, CREOL and Department of Physics, University of Central Florida, 4111 Libra Drive, PS430, Orlando, FL 32816, USA.*
*\*Corresponding author: nariyuki.saito@issp.u-tokyo.ac.jp*

**Photoinduced quantum dynamics in molecules have hierarchical temporal structures with different energy scales that are associated with electron and nuclear motions. Femtosecond-to-attosecond transient absorption spectroscopy (TAS) using high-harmonic generation (HHG) with a photon energy below 300 eV has been a powerful tool to observe such electron and nuclear dynamics in a table-top manner. However, comprehensive measurements of the electronic, vibrational, and rotational molecular dynamics have not yet been achieved. Here, we demonstrate HHG-based TAS at the nitrogen *K*-edge (400 eV) for the first time, and observe all the electronic, vibrational, and rotational degrees of freedom in a nitric oxide molecule at attosecond to sub-picosecond time scales. This method of employing core-to-valence transitions offers an all-optical approach to reveal complete molecular dynamics in photochemical reactions with element and electronic state specificity.**

## 1. INTRODUCTION

The observation of electronic and nuclear dynamics in molecules initiated by light is critical for understanding fundamental mechanisms in photoinduced chemical and physical processes [1]. Pump-probe spectroscopy using high harmonics (HHs) has enabled table-top measurements of ultrafast dynamics with femtosecond-to-attosecond temporal resolution through photoelectron/ion spectroscopy [2-4], high-harmonic spectroscopy [5, 6], and transient absorption spectroscopy (TAS) [7-16]. In particular, TAS has unique advantages over other techniques: it is unaffected by the existence of strong laser fields, state-specific even if the probe pulse has a broadband spectrum, and free of the space charge problem. Because of these characteristics, TAS is one of the most ideal techniques for laser-based pump-probe experiments for a wide range of atoms and molecules. To date, by using TAS in the extreme ultraviolet (XUV) region below ~100 eV, real-time observations of electronic processes such as autoionization [7, 8], valence electron motion [9], tunnel ionization [10], the emergence of laser-dressed states [11] and nuclear dynamics [12-14] have been demonstrated. Couplings between the electronic and nuclear degrees of freedom have also been observed [15, 16]. Recently, thanks to the advances in the development of high-harmonic generation (HHG) driven by long-wavelength, intense infrared (IR) light sources [17-21], TAS in the soft X-ray (SX) region around the water window (from 284 eV to 543 eV) has become possible [22-25]. Short-wavelength SX pulses are expected to enable element-specific TAS in complex molecules in various environments such as in solvents [26] or on surfaces.

However, until now, there has been no TAS measurement that resolves all the electronic, vibrational, and rotational dynamics in molecules simultaneously. Moreover, the maximum photon energy in TAS is mostly limited below 300 eV for molecular targets, which hinders the use of important absorption edges with 1s core states. In particular, TAS in the XUV region has difficulty in resolving molecular rotation because the molecular orbitals that are associated to XUV absorption belong to d or p shells and have relatively complex orbital shapes.

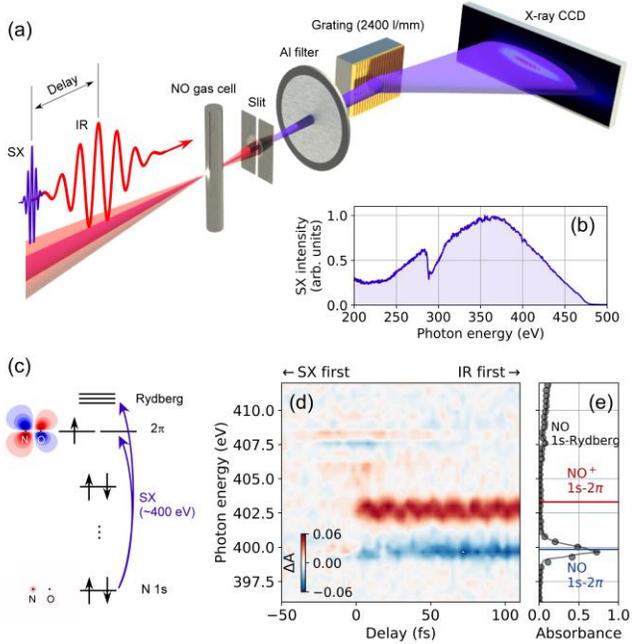

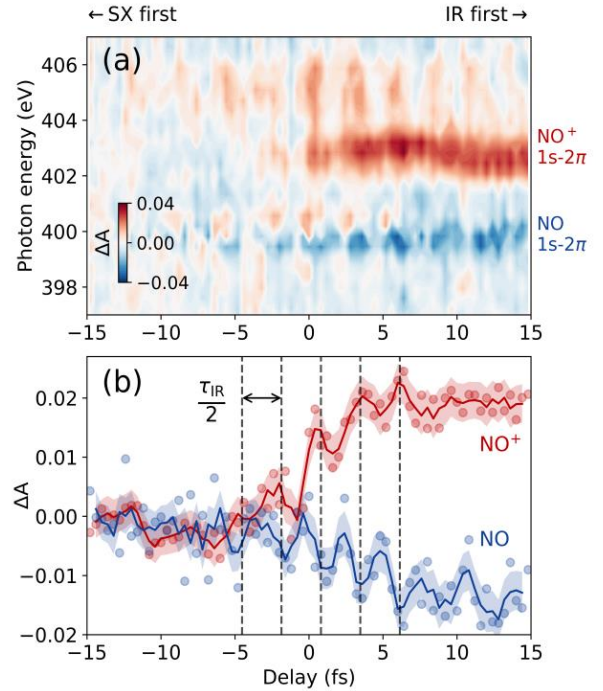

**Fig. 1.** TAS of NO in the SX region. (a) Schematic of the experimental setup. (b) Typical SX spectrum obtained by HHG in helium. (c) Energy levels of NO that are relevant to TAS (right) and the shapes of the N 1s and 2π orbitals (left). (d) Measured transient absorption spectrum. The color scale represents the transient absorbance ΔA (the definition is described in the main text). (e) Static absorbance of NO without the IR pump pulses measured in our experiment (black circles) and in a synchrotron (black curve). The calculated positions of the NO and NO+ 1s-2π peaks are indicated by the blue and red lines, respectively.

Here, we demonstrate that electronic, vibrational, and rotational dynamics at attosecond to sub-picosecond time scales can be simultaneously observed by HHG-based TAS in the SX region. As a target molecule, we choose nitric oxide (NO). NO is not only a good prototype for understanding fundamental dynamics because of its simple structure, but also important in photochemical reactions in the atmosphere or in laser filaments [27]. TAS is conducted with attosecond SX HHs around 400 eV (nitrogen K-edge). Our results will pave the way for the realization of complete measurements of entangled electronic and nuclear dynamics with element specificity.

## 2. RESULTS AND DISCUSSION

### A. TAS of NO Molecules at 400 eV

Figure 1(a) shows a schematic of the experimental setup. We employ a $BiB_3O_6$-based optical parametric chirped pulse amplifier operating at 1.6 μm with a pulse duration of 10 fs as an HHG driver [28]. Attosecond SX pulses obtained by HHG in helium are used as probe pulses while the IR fundamental pulses are used as pump pulses. The two pulses are collinearly focused into an NO gas cell with piezo-controlled delays. The intensity of the pump pulse is set to $\sim 1 \times 10^{14}$ W/cm². The transmitted SX spectra are measured by a spectrometer consisting of a slit, an aluminium filter, a flat-field grating, and an X-ray charge-coupled device (CCD) camera (for details, see Supplementary Section 1). A typical SX spectrum without the NO gas cell is shown in Fig. 1(b). The maximum photon energy reaches above 450 eV, sufficiently covering the nitrogen K-edge. The energy levels of NO relevant to our experiment are depicted in Fig. 1(c). The SX pulse excites an electron in the N 1s core level to the 2π valence level or the Rydberg levels. The measured transient absorption spectrum around 400 eV is shown in Fig. 1(d.) Here, the transient absorbance at a delay of $\tau$ and a photon energy of $E$ is defined as $\Delta A(\tau, E) = -\log_{10}(I(\tau, E)/I_0(E))$, where $I(\tau, E)$ and $I_0(E)$ are the transmitted SX intensities at a delay of $\tau$ and at a sufficiently large

**Fig. 2.** Attosecond strong-field ionization dynamics. (a) Measured attosecond transient absorption spectrum of NO with fine delay steps around the delay origin. (b) Transient absorbance of the NO 1s-2π peak (blue circles, averaged from 398.7 to 400.6 eV) and the NO+ 1s-2π peak (red circles, averaged from 401.7 to 403.6 eV). The solid curves are their 3-point rolling averages, and the shaded areas indicate the errors

negative delay (typically, $\sim$-100 fs), respectively. The measured static absorbance spectrum of NO is shown as black circles in Fig. 1(e). The black curve in Fig. 1(e) is a reference absorbance measured by a synchrotron source [29], which agrees well with our measurement. The assignments of the absorption peaks are obtained by ab initio calculations (for details, see Supplementary Section 2). The strong peak at 399.8 eV and the weak peaks around 407.5 eV are the 1s-2π and 1s-Rydberg peaks of neutral NO, respectively. In the calculation, the position of the 1s-2π peak of NO+ is also determined to be 403.3 eV.

The main feature in the measured transient absorption spectrum in Fig. 1(d) is the suppression of the NO 1s-2π peak and the emergence of the NO+ 1s-2π peak when the IR pulse precedes the SX pulse. This indicates the generation of NO+ by strong-field ionization. Moreover, around the delay origin, the NO 1s-Rydberg peaks are modulated in the presence of a strong IR field. This is due to the AC Stark shift and the emergence of laser-dressed states, which are similar to those observed in previous TAS studies of atoms and molecules [11, 15, 16, 24]. Here, we focus on the NO and NO+ 1s-2π peaks and discuss the dynamics of strong-field ionization and the subsequent nuclear motions.

### B. Attosecond Electronic Dynamics

First, we describe the attosecond strong-field ionization dynamics. Figure 2(a) shows the transient absorption spectrum with fine delay steps (400 as) around the delay zero. The line plots of the transient absorbance of the NO and NO+ 1s-2π peaks are shown in Fig. 2(b). In both NO and NO+ 1s-2π peaks, in addition to the monotonic decrease and increase of the absorbance due to ionization, clear oscillatory structures with a period of a half cycle of the pump IR pulse (2.7 fs) are observed. Moreover, the phase of the oscillation is opposite for the NO and NO+ 1s-2π peaks, indicating that the possible origin of the oscillation is the change of the hole population in NO during strong-field ionization. This type of oscillation was recently observed in xenon atoms by Sabbar

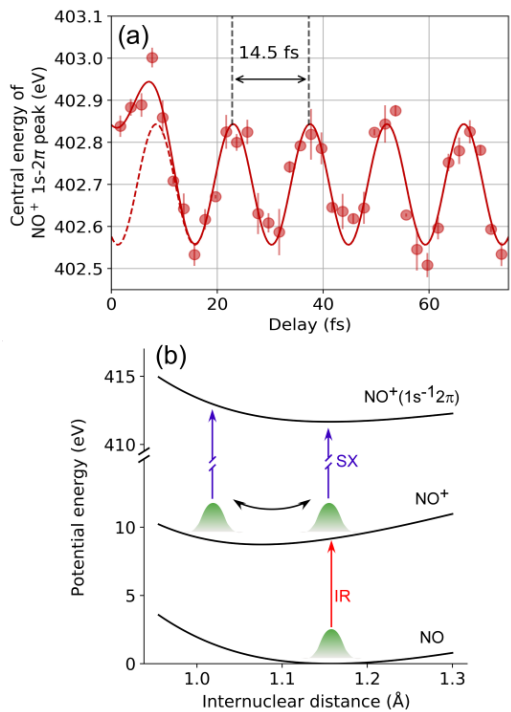

**Fig. 3.** Coherent molecular vibration. (a) Measured central energy of the NO$^+$ 1s-2π peak as a function of the delay (circles). The solid curve is the fitting result with a cosine function plus a Gaussian function around the time origin. The dashed curve is the cosine component of the fitting result. (b) Potential energy curves of the relevant states obtained by the ab initio calculation and a schematic of the mechanism of the molecular vibration.

*et al.* [30], and called the ground-state polarization effect. In typical tunnel ionization, the hole population increases stepwise, reflecting the fact that tunnel ionization is a highly nonlinear process and that ionization dominantly occurs at the local maxima of the laser intensity [3, 10]. However, when ground-state polarization takes place in addition to tunnel ionization, a fraction of the outgoing electron wave packet is pulled back to the parent ion by its Coulomb potential, resulting in an oscillatory structure in the temporal evolution of the hole population. The clear oscillation observed in our experiment will provide a clue to understand the electronic dynamics in molecules which occur at the sub-optical-cycle time scale during strong-field ionization.

### C. Femtosecond Vibrational Dynamics

Second, we discuss the vibrational dynamics of the electronic ground state of NO$^+$ triggered by strong-field ionization. Figure 3(a) shows the delay-dependent central energies of the NO$^+$ 1s-2π peak extracted by Gaussian fitting. A clear oscillation with a period of 14.5±0.1 fs is observed. The obtained oscillation period agrees well with the value from the literature for the vibration period of NO$^+$ (14.23 fs) [31]. The initial phase of the vibration is determined to be (0.17±0.08) π rad, where 0π and 0.5π mean perfect "cosine-like" and "sine-like" phases, respectively. This implies that the observed oscillation is close to "cosine-like."

The mechanism of the observed oscillation in the 1s-2π peak energy, as well as its initial phase, can be explained by the potential energy curves of NO and NO$^+$ obtained by the ab initio calculation (Fig. 3(b)). Upon ionization, a vibrational wave packet is created on the electronic ground state of NO$^+$. Because the equilibrium internuclear distance of NO$^+$ is smaller than that of NO, the created vibrational wave packet in NO$^+$ is displaced from the bottom of the potential energy curve and thus starts vibrating. The oscillation of the internuclear distance is directly mapped to the transition energy between the 1s and 2π levels, as the transition energy monotonically decreases with respect to the internuclear distance. This mechanism [13, 14] is consistent with the experimentally observed "cosine-like" initial phase. In addition to the coherent vibration, the observed central energy of the NO$^+$ 1s-2π peak contains a unidirectional upper shift around the delay origin, which might originate from field-induced effects.

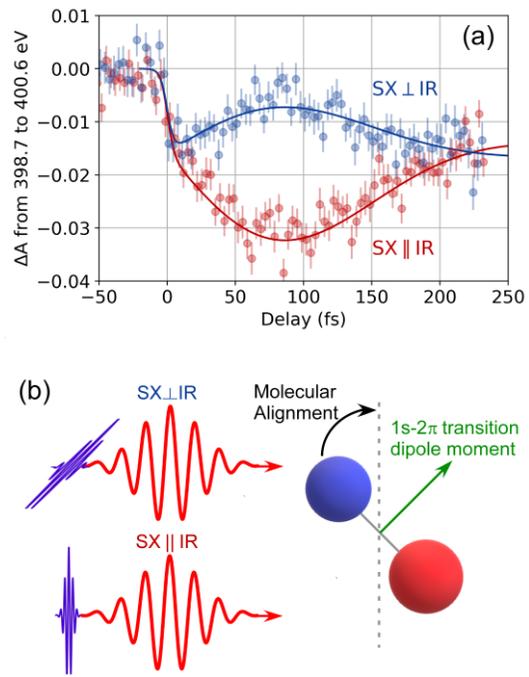

**Fig. 4.** Coherent molecular rotation. (a) Measured (circles) and calculated (solid curves) transient absorbance of the NO 1s-2π peak averaged from 398.7 to 400.6 eV. The blue and red data are measured when the polarizations of the SX and IR pulses are perpendicular and parallel, respectively. (b) Schematic of the mechanism of the SX absorption change upon molecular alignment.

### D. Sub-picosecond Rotational Dynamics

Finally, we describe the rotational dynamics. Figure 4(a) shows the transient absorbance of the NO 1s-2π peak, averaged from 398.7 to 400.6 eV. The circles are the measured data, and the solid curves are the calculation results. The calculation contains two signals: tunnel ionization and molecular rotation. The tunnel ionization signal is the abrupt decrease of the absorbance around the delay zero, whose shape is determined from sigmoid fitting of the transient absorbance of the NO$^+$ 1s-2π peak. The molecular rotation signal is the slow increase or decrease of the absorbance, which is simulated by the time-dependent Schrödinger equation for the rotational states of NO with a pump intensity of 1.3×10$^{14}$ W/cm$^2$ (for details, see Supplementary Section 3). The experiment and the calculation are in good agreement.

The mechanism of the molecular rotation and its effect on the SX absorption are depicted in Fig. 4(b). When an NO molecule is irradiated by a linearly polarized strong IR pulse, it begins to align with the polarization of the IR pulse via the interaction between the induced dipole moment and the laser electric field [32]. Meanwhile, the transition dipole moment between the 1s and 2π orbitals is perpendicular to the molecular axis. Therefore, if the polarization of the SX pulse is perpendicular to that of the IR pulse, the inner product between the SX polarization and the 1s-2π transition dipole moment increases as the NO molecule aligns, resulting in an increase in the SX absorption. Conversely, if the polarization of the SX pulse is parallel to that of the IR pulse, the SX absorption decreases. In Fig. 4(a), there is a bump structure with a peak at approximately 80 fs. This reflects the fact that the molecular alignment occurs at ~80 fs; subsequently, the alignment is lost due to the dephasing among the rotational states.

Note that the rotational dynamics observed in our experiment are difficult to access by conventional TAS in the XUV region using d or p inner shells. The reason is that the d or p inner shells are usually energetically degenerate and the transition dipole moment vector for each degenerate state is oriented differently.

## 3. CONCLUSION

In summary, we demonstrate that electronic and nuclear dynamics of diatomic molecules at attosecond to sub-picosecond time scales can simultaneously be observed by TAS in the SX region, around 400 eV, in a table-top manner. The target sample in our experiment is a prototypical diatomic molecule, NO, but the same technique can be applied to more complex molecules in the gas phase or in solvents, where the element specificity of SX absorption can be fully utilized to understand the molecular dynamics that occur in multiple degrees of freedom at different time scales. Such a technique will be generally applicable to measure the couplings among electronic, vibrational, and rotational dynamics, which are quite common in strong field-induced molecular dynamics [33, 34] and in photo-induced phase-transition dynamics in charge-transfer complexes [35]. Another possible application is photocatalysis, where electronic charge transfer and nuclear dynamics (e.g., molecular vibration and molecular orientation with respect to the catalyst surface) play critical roles [36, 37]. Attosecond TAS in the SX region can possibly resolve these couplings in transient states during chemical reactions or phase transitions, which is difficult to access with other techniques.

**Funding.** JSPS KAKENHI Grant Numbers JP18H05250, JP18J11286; MEXT Q-LEAP Grant Number JPMXS0118068681; Advanced Leading Graduate Course for Photon Science (ALPS); Air Force Office of Scientific Research (AFOSR) (FA9550-16-1-0013); Army Research Office (ARO) (W911NF-14-1-0383); Defense Advanced Research Projects Agency (DARPA) (D18AC00011); National Science Foundation (NSF) (1806584).

**Acknowledgment**. The authors would like to thank Prof. Akiyoshi Hishikawa, Dr. Mizuho Fushitani, Dr. Akitaka Matsuda, Dr. Hirofumi Yanagisawa, and Mr. Kazma Komatsu for fruitful discussions.

**Disclosures**. The authors declare that there are no conflicts of interest related to this article.

See Supplement 1 for supporting content.

# Real-time observation of electronic, vibrational, and rotational dynamics in nitric oxide with attosecond soft X-ray pulses at 400 eV: supplementary material

**NARIYUKI SAITO,**[1,*] **HIROKI SANNOHE,**[1] **NOBUHISA ISHII,**[2] **TERUTO KANAI,**[1] **NOBUHIRO KOSUGI,**[3] **YI WU,**[4] **ANDREW CHEW,**[4] **SEUNGHWOI HAN,**[4] **ZENGHU CHANG,**[4] **AND JIRO ITATANI**[1]

[1]*The Institute for Solid State Physics, the University of Tokyo, Kashiwanoha 5-1-5, Kashiwa, Chiba 277-8581, Japan.*
[2]*Kansai Photon Science Institute, National Institutes for Quantum and Radiological Science and Technology, 8-1-7 Umemidai, Kizugawa, Kyoto 619-0215, Japan*
[3]*Institute of Materials Structure Science, High Energy Accelerator Research Organization, 1-1 Oho, Tsukuba, Ibaraki 305-0801, Japan.*
[4]*Institute for the Frontier of Attosecond Science and Technology, CREOL and Department of Physics, University of Central Florida, 4111 Libra Drive, PS430, Orlando, FL 32816, USA.*
*Corresponding author: nariyuki.saito@issp.u-tokyo.ac.jp*

This document provides supplementary information to "Real-time observation of electronic, vibrational, and rotational dynamics in nitric oxide with attosecond soft X-ray pulses at 400 eV."

## 1. EXPERIMENTAL SETUP

A detailed schematic of the TAS beamline is presented in Fig. S1. The IR pulses obtained from the $BiB_3O_6$-based optical parametric chirped-pulse amplifier (1.6 μm, 10 fs, 1.5 mJ, 1 kHz) are split into pump and probe arms by a beam splitter. The pump and probe arms contain 10% and 90% of the total pulse energy, respectively. In the probe arm, the IR pulses are focused by a lens (f = 50 cm) into a semi-infinite helium gas cell (2.4 bar) to generate SX HHs. A two-stage differential pumping system after the gas cell using two dry pumps (500 l/min.) reduces the gas pressure in the subsequent vacuum chambers in the beamline. The SX pulses are passed through an aluminium filter (150 nm) to remove the fundamental IR component and focused into an NO gas cell (0.1 bar, 1.5 mm thick) by a toroidal mirror (4f = 2 m). The SX pulse duration is estimated to be ~200 as by SFA calculation. The pump IR pulses are recombined with the probe SX pulses by a hole-drilled mirror and collinearly focused into the NO gas cell by a lens (f = 25 cm). The IR intensity is estimated to be ~$1 \times 10^{14}$ W/cm². The delay between the SX and IR pulses is scanned by a piezo stage. In the attosecond TAS measurement (Fig. 2 in the main text), the delay is controlled to a precision of ~30 as with a feedback system using a HeNe laser which propagates collinearly with the SX and IR pulses [S1]. The SX spectra are recorded by a spectrometer consisting of a slit (50 μm), a flat-field grating (Shimadzu, 2400 l/mm), and a back-illuminated X-ray CCD camera (Andor, Newton SO). The photon flux at 400 eV at the CCD is ~80 photons/s/eV.

The carrier-envelope phase (CEP) of the IR pulses is passively stabilized during attosecond TAS. The CEP-dependence of the HH spectra is shown in Fig. S2(a), showing clear half-cycle cut-off structures [S2]. In attosecond TAS, an HH spectrum at a CEP of 0.8π in Fig. S2(a) is employed (Fig. S2(b)). The nitrogen *K*-edge (400 eV) is covered by only one half-cycle cutoff, indicating that we use isolated attosecond SX pulses. In the other measurements, the CEP is rapidly scanned from shot to shot with a step of 0.1π by using an acousto-optic programmable dispersive filter in the laser system. Practically, this corresponds to a "randomization" of the CEP. The

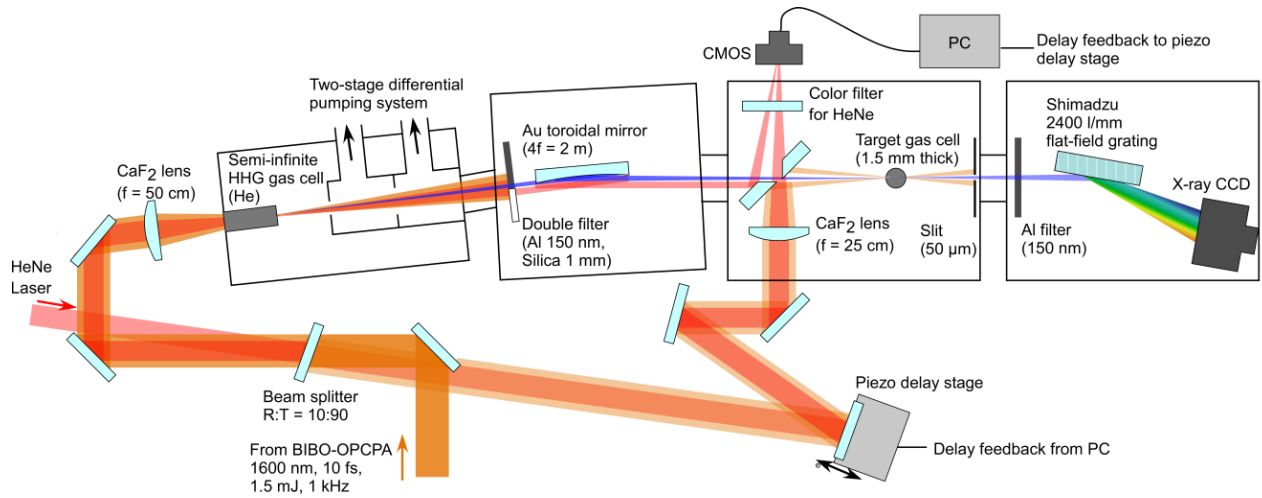

**Fig. S1.** Detailed schematic of the TAS beamline. The delay feedback system is employed only in attosecond TAS.

HH spectrum shown in Fig. 1(b) in the main text is measured under such a condition. The randomization of CEP reduces the fine spectral structures in the HH spectrum, resulting in a better signal-to-noise ratio in TAS.

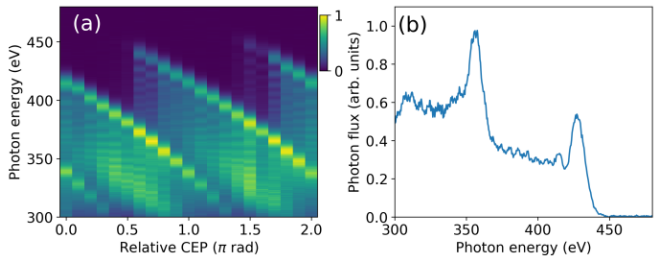

**Fig. S2.** CEP-dependent HH spectra. (a) Measured HH spectra with the relative CEP scanned from 0 to $2\pi$. (b) HH spectrum employed in attosecond TAS (CEP = $0.8\pi$ in (a)).

## 2. AB INITIO CALCULATION OF THE GROUND STATES AND THE CORE-EXCITED STATES OF NO

The neutral and cationic ground states, NO ($^2\Pi$) and NO$^+$ ($^1\Sigma^+$), and the lowest N 1s-2$\pi$ ($\pi^*$) excited states, NO*($^2\Delta$) and NO$^{+*}$($^1\Pi$), were obtained by configuration-interaction (CI) calculations including single and double substitutions up to 17 virtual orbitals from the full valence configuration space over 3-6 $\sigma$ and 1-2 $\pi$ orbitals using the self-consistent field (SCF) orbitals for each state, where the core orbitals are frozen and their occupation numbers are fixed in the CI calculations. Primitive basis functions were taken from (73/7) contracted Gaussian-type functions [S3]. They were augmented with two d-type polarization functions for each atom: the exponential $\zeta = 2.704$ and $0.535$ for oxygen, $\zeta = 1.986$ and $0.412$ for nitrogen. The contraction schemes were (4111111/31111/1*1*). The calculations were performed by using the originally developed quantum chemical computation code GSCF3 [S4, S5]. The results are shown in Table S1. We should note the second and third lowest N 1s-$\pi^*$ states, NO*($^2\Sigma^-$) and NO*($^2\Sigma^+$), contribute to the observed N 1s-$\pi^*$ band [29], though we do not use these state energies in the present work.

**Table S1.** Total energies $E_{total}$ in a.u. (1 a.u. = 27.2116 eV) obtained from the CI calculations for the neutral and cationic ground states, NO ($^2\Pi$) and NO$^+$ ($^1\Sigma^+$), and the lowest N 1s-2$\pi$ excited states, NO*($^2\Delta$) and NO$^{+*}$($^1\Pi$).

| Internuclear distance (Å) | Ground states | | Lowest N 1s-2$\pi$ excited states | |
|---|---|---|---|---|
| | NO ($^2\Pi$) | NO$^+$ ($^1\Sigma^+$) | NO* ($^2\Delta$) | NO$^{+*}$ ($^1\Pi$) |
| 0.95 | -129.3957 | -129.1544 | -114.6281 | -114.2782 |
| 1.05 | -129.5055 | -129.2120 | -114.7805 | -114.3805 |
| 1.15 | -129.5346 | -129.2007 | -114.8421 | -114.4061 |
| 1.25 | -129.5213 | -129.1582 | -114.8551 | -114.3952 |
| 1.35 | -129.4871 | -129.1040 | -114.8435 | -114.3722 |

## 3. CALCULATION OF MOLECULAR ROTATION AND ITS EFFECT ON SX ABSORPTION

Laser-induced molecular rotation is simulated by solving the time-dependent Schödinger equation for the rotational states of NO, as described in [32]. The polarizability and rotational constant required for the calculation are taken from [S6, S7]. As an excitation IR pulse, we use a Gaussian pulse (1.6 µm, 10 fs FWHM, $1.3 \times 10^{14}$ W/cm$^2$) that is similar to those employed in the experiment.

The degree of molecular alignment can be evaluated by $<\cos^2\theta>$, where $\theta$ is the angle between the IR polarization vector and the molecular axis and $<\cdot>$ means a thermally averaged expectation value. The calculated $<\cos^2\theta>$ as a function of time is plotted in Fig. S3 for various temperatures. Note that the temperature of NO is 300 K in the actual measurement because NO is filled in a gas cell.

The relationship between molecular alignment and SX absorption is formulated as follows. We assume that the transition dipole moment between the inner shell and the valence orbital is perpendicular to the molecular axis. If the polarizations of the IR and SX pulses are parallel, the inner product between the transition dipole moment and the SX polarization vector is proportional to $\sin\theta$. Therefore, the SX absorption is proportional to $<\sin^2\theta> = 1 - <\cos^2\theta>$. On the other hand, if the polarizations of the IR and SX pulses are perpendicular, the inner product is proportional to $\sqrt{1 - \sin^2\theta \cos^2\varphi}$, where $\varphi$ is the polar angle. In this case, the SX absorption is proportional to $<\int(1-\sin^2\theta\cos^2\varphi)d\varphi/2\pi> = (1+<\cos^2\theta>)/2$. By using these expressions, the relative SX absorption change can be evaluated from the calculated degree of alignment, <cos²θ>.

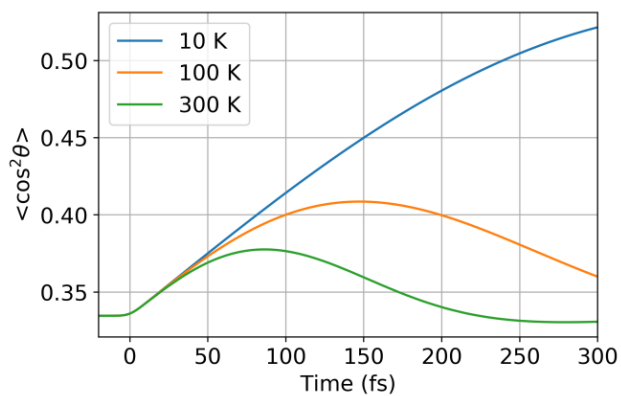

**Fig. S3.** Calculated degree of alignment of NO irradiated by a 1.6 μm, 10 fs, 1.3×10$^{14}$ W/cm$^2$ Gaussian pulse for various rotational temperatures.